\documentclass{PoS}
\usepackage{bbm}
\usepackage[utf8]{inputenc}

\title{Meson masses in external magnetic fields with HISQ fermions}

\ShortTitle{Meson masses in external magnetic fields with HISQ fermions}

\author{Heng-Tong Ding$^1$, Sheng-Tai Li $^{2,1}$, Swagato Mukherjee$^3$, Akio Tomiya$^4$, \speaker{Xiao-Dan Wang}$^1$\thanks{This work was supported by the National Natural Science Foundation of China under grant numbers 11535012, 11775096, 11947237 and Huabo fellowship in Central China Normal University (Xiao-Dan Wang) and the RIKEN Special Postdoctoral Researcher program (Akio Tomiya). This work was also supported by the U.S. Department of Energy, Office of Science, Office of Nuclear Physics through the Contract No. DE-SC0012704 and the U.S. Department of Energy, Office of Science, Office of Nuclear Physics and Office of Advanced Scientific Computing Research within the framework of Scientific Discovery through Advance Computing (SciDAC) award Computing the Properties of Matter with Leadership Computing Resources. The numerical simulations have been performed on the GPU cluster in the Nuclear Science Computing Center at Central China Normal University (NSC$^3$), Wuhan, China.}\\
       $^1$ Key Laboratory of Quark \& Lepton Physics (MOE) and Institute of Particle Physics, Central China Normal University, Wuhan 430079, China\\
       $^2$ Institute of Modern Physics,Chinese Academy of Sciences, Lanzhou 730000, China \\
       $^3$ Physics Department, Brookhaven National Laboratory, Upton, NY 11973, USA\\
       $^4$ RIKEN/BNL Research center, Brookhaven National Laboratory, Upton, NY, 11973, USA\\
        E-mail: \email{hengtong.ding@mail.ccnu.edu.cn,lishengtai@mails.ccnu.edu.cn, swagato@bnl.gov, akio.tomiya@riken.jp, xiaodanwang@mails.ccnu.edu.cn}}

\abstract{We studied the temporal correlation function of mesons in the pseudo-scalar channel in (2+1)-flavor QCD in the presence of external magnetic fields at zero temperature. The simulations were performed on $32^3 \times 96$ lattices using the Highly Improved Staggered Quarks (HISQ) action with $m_{\pi} \approx $  230 MeV. The strength of magnetic fields $|eB|$ ranges from 0 to around 3.3 GeV$^2$ ($\sim 60 m_\pi^2$). We found that the masses of neutral pseudo-scalar particles, e.g. neutral pion and kaon, monotonouslly decrease as the magnetic field grows and then saturate at a nonzero value. It is observed that heavier neutral pseudo-scalars are less affected by magnetic fields. Moreover, we found a non-monotonous behavior of  charged pion and kaon mass in magnetic field for the first time. In the case of small magnetic field (0 $\leq~|eB| \lesssim$ 0.3 GeV$^2~\sim 6m_\pi^2$ )  the mass of charged pseudo-scalar grows with magnetic field and can be well described by the Lowest Landau Level approximation, while for $|eB|$ larger than 0.3 GeV$^2$ the mass starts to decrease. The possible connection between $|eB|$ dependences of neutral pion mass and the decreasing behavior of pseudo-critical temperature in magnetic field is discussed. Due to the nonzero value of neutral pion mass our simulation indicates that the superconducting phase of QCD does not exist in the current window of magnetic field. }

\FullConference{37th International Symposium on Lattice Field Theory - Lattice2019\\
		16-22 June 2019\\
		Wuhan, China}

\begin{document}

\section{Introduction}
Strong magnetic fields have significant impacts on the dynamics of strong interaction. Such strong magnetic can be produced in heavy-ion collision \cite{Kharzeev:2007jp,Skokov:2009qp,Deng:2012pc}, early Universe \cite{Vachaspati:1991nm} and magnetars \cite{enqvist1993primordial}.
Many interesting effects of QCD in external magnetic field have been found in recent studies, and one of them is the so-called inverse magnetic catalysis (IMC) \cite{Bali:2012zg}, which is a non-monotonous behavior of the condensate as a function of magnetic field $|eB|$ near transition temperature $T_{pc}$ observed by performing a full QCD simulation. The IMC accompanies with a decreasing of transition temperature $T_{pc}$ with increasing magnetic field \cite{Bali:2011qj}. These two phenomena are not expected from early studies by effective models. Investigating meson spectrum of QCD, in particular neutral pion in external magnetic field may thus be helpful to understand the behavior of $T_{pc}$ assuming  that neutral pion is still a Goldstone boson in the nonzero magnetic field. 

The $|eB|$ dependence of meson masses is also relevant in the study of phase structure of QCD in presence of background magnetic field. Extended Nambu-Jona-Lasinio (NJL) model calculation suggests that a sufficiently strong magnetic field may turn QCD vacuum into a superconducting state \cite{Chernodub:2010qx,Chernodub:2011mc}. This is signaled by the charged $\rho$ meson condensation. QCD inequalities \cite{Hidaka:2012mz} imply that the vector channel meson correlation function is bounded by the connected neutral pion correlation function, therefore their masses are also bounded. This indicates that neutral pion mass must vanish if $\rho$ meson condensates.

Most of studies on meson spectrum in external magnetic fields have been performed in quenched QCD, e.g. quenched two color QCD with overlap fermions as valence quarks \cite{luschevskaya2014rho}, quenched  QCD with Wilson fermions \cite{Bali:2017ian,Hidaka:2012mz} and overlap fermions \cite{luschevskaya2015magnetic} as valence quarks.  Early studies found that the masses of neutral light pseudo scalars firstly decrease and then increase as magnetic field grows~\cite{Hidaka:2012mz} while the mass of charged light pseudo scalar ($\pi^{\pm}$) increases as magnetic field increases. Later it was pointed out in Ref.\cite{Bali:2017ian} that the $|eB|$-dependence of hopping parameter $\kappa$ has to be taken into consideration. Corrected with addictive mass renormalization to determine $\kappa$, a monotonous reduction of neutral pion mass and a monotonous increasing of charged pion mass as magnetic field grows are finally observed\cite{Bali:2017ian}.
On the other hand, investigations in full QCD are restrained for light pseudo scalar particles ($\pi_0,\pi^{\pm}$)~\cite{Bali:2011qj,Bali:2017ian}.  It is shown in Ref. ~\cite{Bali:2011qj,Bali:2017ian} that behavior of masses of $\pi_0$ and $\pi^{\pm}$ in external magnetic fields obtained from (2+1)-flavor QCD using stout fermions have similar trends as those from quenched QCD.

In this proceedings we will report our studies on light and strange pseudo scalar mesons in $N_f=2+1$ QCD using HISQ (Highly Improved Staggered Quark) action with the strength of magnetic field up to around $60m_\pi^2$.  We will give simulation details in Section~\ref{sec:setup}, and present results in Section~\ref{sec:results} and summarize in Section~\ref{sec:conclusion}.
  
\section{Lattice Setup and Methodology}
\label{sec:setup}
We employed $N_f=2+1$ HISQ action and tree-level improved Symanzik gauge action on $32^3\times 96$ lattice with $m_\pi$ around 230 MeV at zero temperature. The external magnetic field is applied along z-direction $\mathbf{B}=(0,0,B)$, and quantized in the following way 

\begin{equation}
q B= \frac{2 \pi N_{b}}{N_{x} N_{y}}a^{-2}.
\end{equation} 
Here $a$ is the lattice spacing with $a^{-1}=1.6852$ GeV.  Taking into account that the difference between u,d,s quark's charge ($\frac{2}{3}e,-\frac{1}{3}e,-\frac{1}{3}e$), a common divisor $\frac{1}{3}$ has been chosen in the quantization $eB= \frac{6 \pi N_{b}}{N_{x} N_{y}} a^{-2}$. $N_b \in \mathbf{Z}$ is the number of magnetic flux through unit area in the x-y plane which is limited by the boundary condition ($0<N_b<\frac{N_xN_y}{4}$). In our simulations $N_b$ is chosen to be 0, 1, 2, 3, 4, 6, 8, 10, 12, 16, 20, 24, 32, 48, 64 whose corresponding magnetic strength ranges from 0 to around 3 GeV$^2$. All configurations have been produced using Rational Hybrid Monte Carlo (RHMC) algorithm and are saved by every 5 time units, statistic for each $N_b$ is about $O(\sim10^3)$.

\begin{table}[htp]
\begin{center}
\begin{tabular}{c|c|c|c|c|c|c|c|c}
\hline
$|eB|$(GeV$^2$) & 0 & 0.052 & 0.104 & 0.157 & 0.209 & 0.314 & 0.418 & 0.523   \\
\hline
$N_b$ & 0 & 1 & 2 & 3 & 4 & 6 & 8 & 10 \\
\hline
\# of conf. & 2548 & 2651 & 3135 & 2444 & 2224 & 3142 & 2935 & 3302 \\
\hline

\hline
$|eB|$(GeV$^2$) & 0.627 & 0.836 &  1.045  & 1.255 & 1.673 & 2.510 & 3.345    \\
\hline
$N_b$ & 12 & 16 & 20 & 24 & 32 & 48 & 64 \\
\hline
\# of conf. & 3432 & 1993 &  3160 & 1994 & 2174 & 995 & 356  \\
\hline
\end{tabular}
\end{center}
\caption{The statistics of measured correlators at different values of $N_b$ }
\label{tab.statistics}
\end{table}

We measured temporal correlation function of staggered bilinear according to the formula below\footnote{The disconnected part in the neutral pion correlation function is neglected. And ${\pi^0_u}$ (${\pi^0_d}$) labels quantities obtained from the up(down) quark component of the correlation function. }
\begin{equation}
\mathrm{G}(\tau)=-\sum_{n} \zeta(\vec{n}) \mathrm{Tr} \left[M^{-1 \dagger}(n, \tau) M^{-1}(n, 0)\right],
\end{equation}
where $\zeta(\vec{n})$ is the phase factor characterizing Dirac and flavor indices of staggered fermions. In this work, we focuse on the correlators in pseudo-scalar channel. Take $\pi^+ (\overline{d}\gamma_5u)$ as an example, combining $\gamma_5$ with staggered fermion Dirac structure and $\gamma_5$-hermiticity gives the phase factor $\zeta(\vec{n})=1$ for charged pion correlation function $G(\tau)_{\pi^+}$. 
As shown in the left plot of Fig. 1, the correlators increase with increasing $N_{b}$ values, which means that the decreasing of corresponding particle mass is expected.

\begin{figure}[htbp]
\centering
\includegraphics[width=0.45\textwidth]{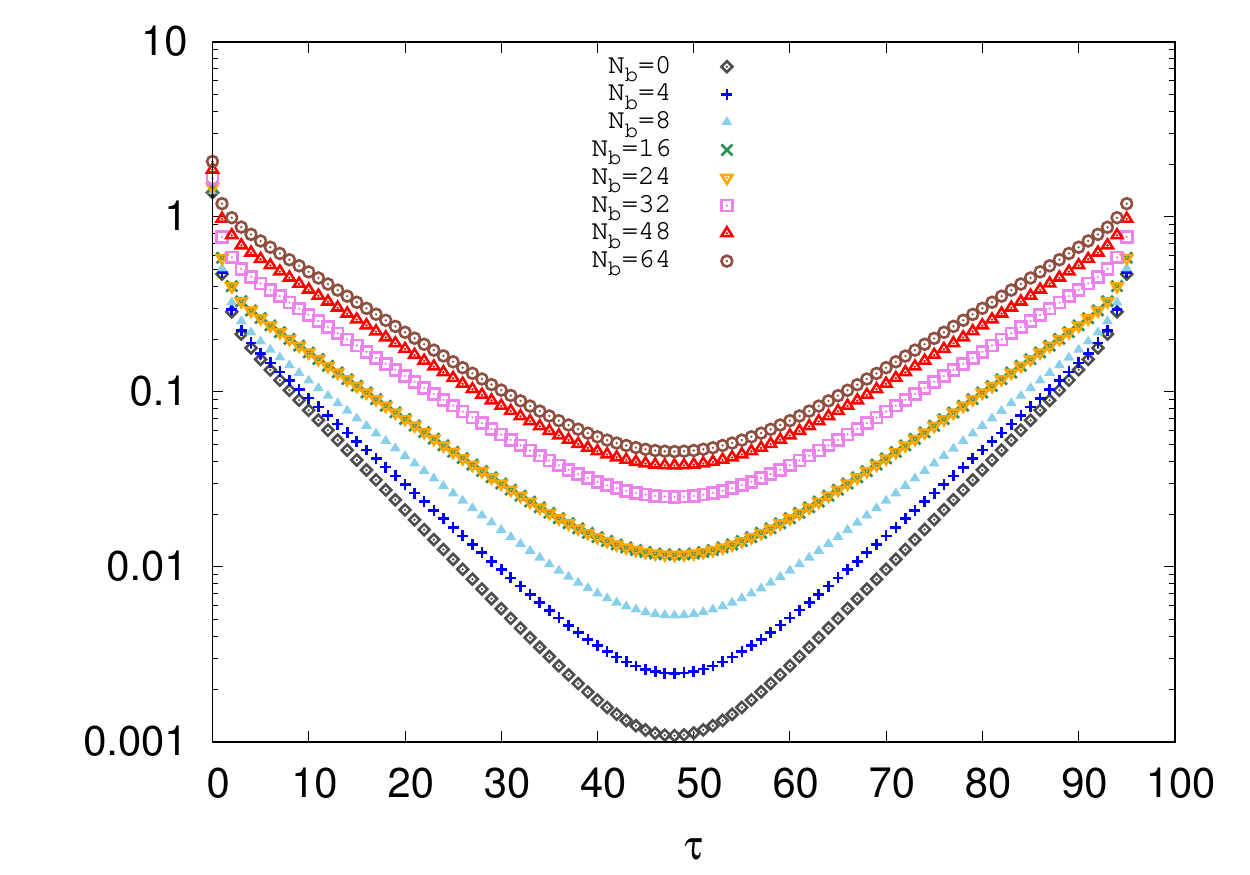}
\includegraphics[width=0.45\textwidth]{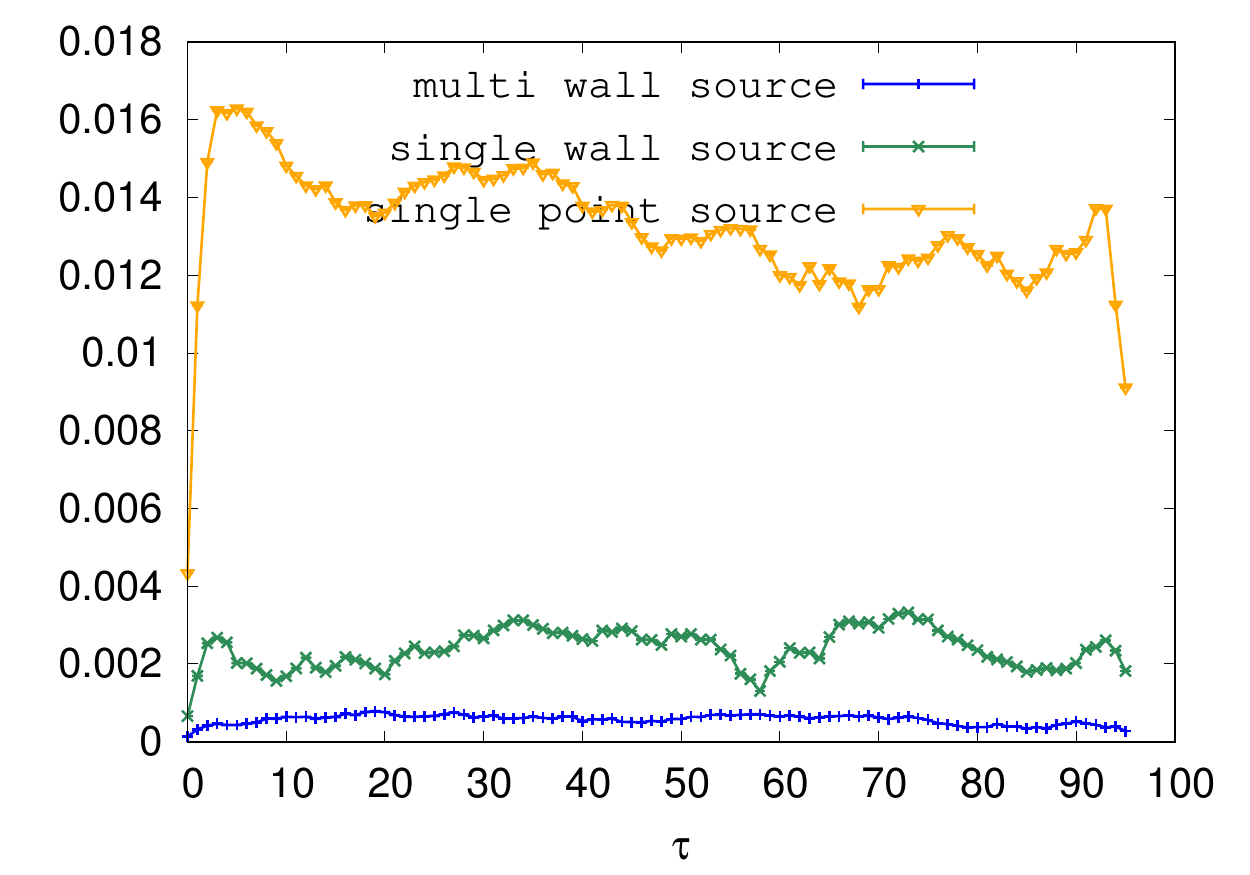}
\caption{Left: An example of correlation function $G(\tau)_{\pi^0_u}$:
 at different $N_b$ values calculated with point source. Right: Comparison of the signal to noise ratio (error/mean values) obtained using different sources.}\label{fig.corr}
\end{figure}

The multiple corner wall sources were used to improve the signal of ground state. As shown in the right plot of Fig. \ref{fig.corr}, the signal to noice ratio $\delta G(\tau) / \langle G(\tau)\rangle$ obtained using the corner wall source is reduced by a factor of 6 compared to the results obtained using point source. We put 12 corner wall sources at (0,0,0,0), (0,0,0,8),...,(0,0,0,88) to further improve the signal. The signal to noise ratio obtained using multiple corner wall source is found $\sqrt{\#\ \mathrm{of\ sources}}$ times better than using single corner wall sources.

The effective mass can be obtained by solving the cosh equation. But for the case of charged pseudo-scalar particles, the signals are too noisy to extract effective mass in this way. In our study, meson masses are obtained by fitting the correlators according to the following ansatz,

\begin{equation}
G_{\mathrm{fit}}\left(n_{\tau}\right)=\sum_{i=1}^{N_{s}, n o} A_{n o, i} \exp \left(-m_{n o, i} n_{\tau}\right)-(-1)^{n_{\tau}} \sum_{i=0}^{N_{s}, o s c} A_{o s c, i} \exp \left(-m_{o s c, i} n_{\tau}\right),
\end{equation}
where $N_{s, no}$ ( $N_{s, osc}$) is the number of non-oscillating (oscillating) states we chose to fit to  the correlators. Non-oscillating states are the physical states we need, but oscillating states are also necessary in fitting correlators of charged particles . Fit modes ($N_{s, no}$\ ,\ $N_{s, osc}$) have been set to (1,0), (1,1) and (2,1). 

 \begin{figure}[htbp]
 \centering
\includegraphics[width=0.6\textwidth]{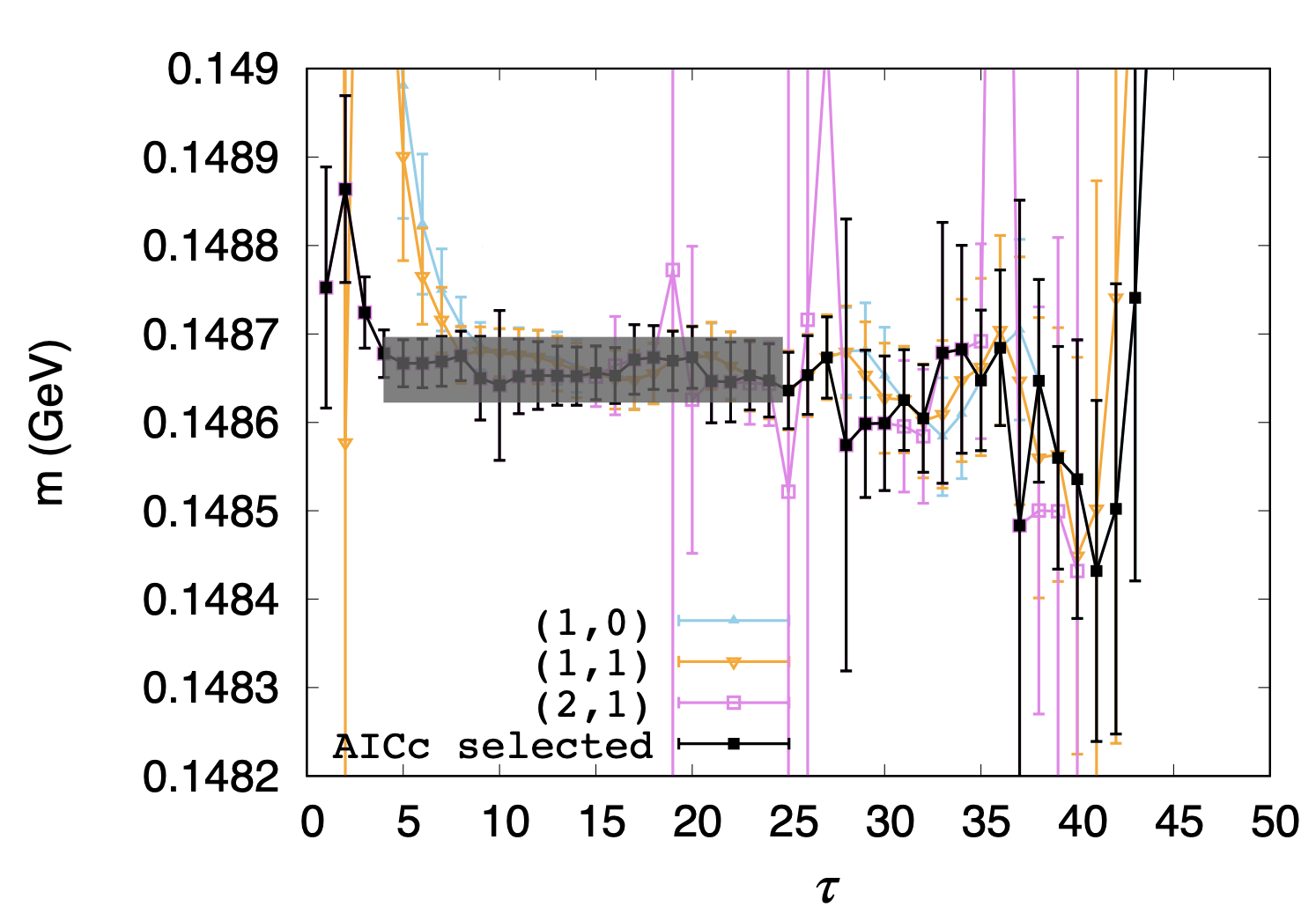} 
\caption{Mass of $\pi^0_u$ at $N_b=16$ ($|eB|$ = 0.84 GeV$^2$) with there different fit modes: (\# of non-oscillating states, \# of oscillating states) = (1,0), (1,1), (2,1) . Black points at each $\tau$ are the best final state chosen by AICc.}\label{fig.aicc}
\end{figure}

The first step to arrive at the final results is to choose the best fit from all different fit modes through AICc (corrected Akaike Information criterion) \cite{1100705, cavanaugh1997unifying}
\begin{equation}
\mathrm{AICc}=2 k-\ln (\hat{L})+\mathrm{AIC}+\frac{2 k^{2}+2 k}{n-k-1},
\end{equation}
where $k$ is the number of parameters, $\hat{L}$ is the likelihood function, and the last term is needed to correct over-fitting if the number of data points $n$ is not much larger than $k$.
Then we choose a plateau in the AICc selected results and obtain the final mass and its uncertainty by using Gaussian bootstrapping method from the plateau.
\section{Results}
\label{sec:results}
Here we present our study on the $|eB|$ dependences of neutral pseudo-scalar particles ($\pi^0_u$, $\pi^0_d$, $K$, $\eta_s$) and charged pseudo-scalar particles ($\pi^-$ and $K^-$).

\begin{figure}[htbp]\label{fig.neutral}
\centering
\includegraphics[width=0.48\textwidth]{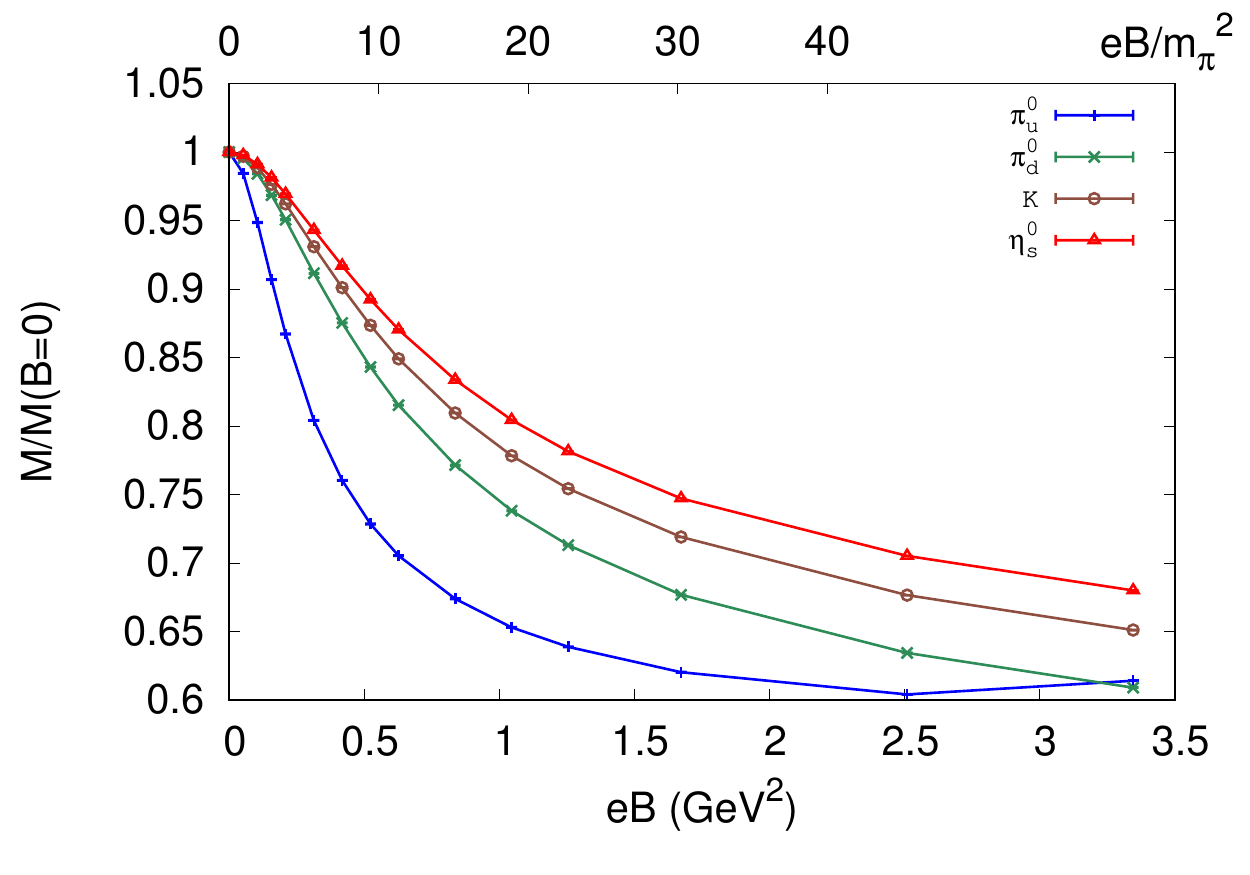}
\includegraphics[width=0.48\textwidth]{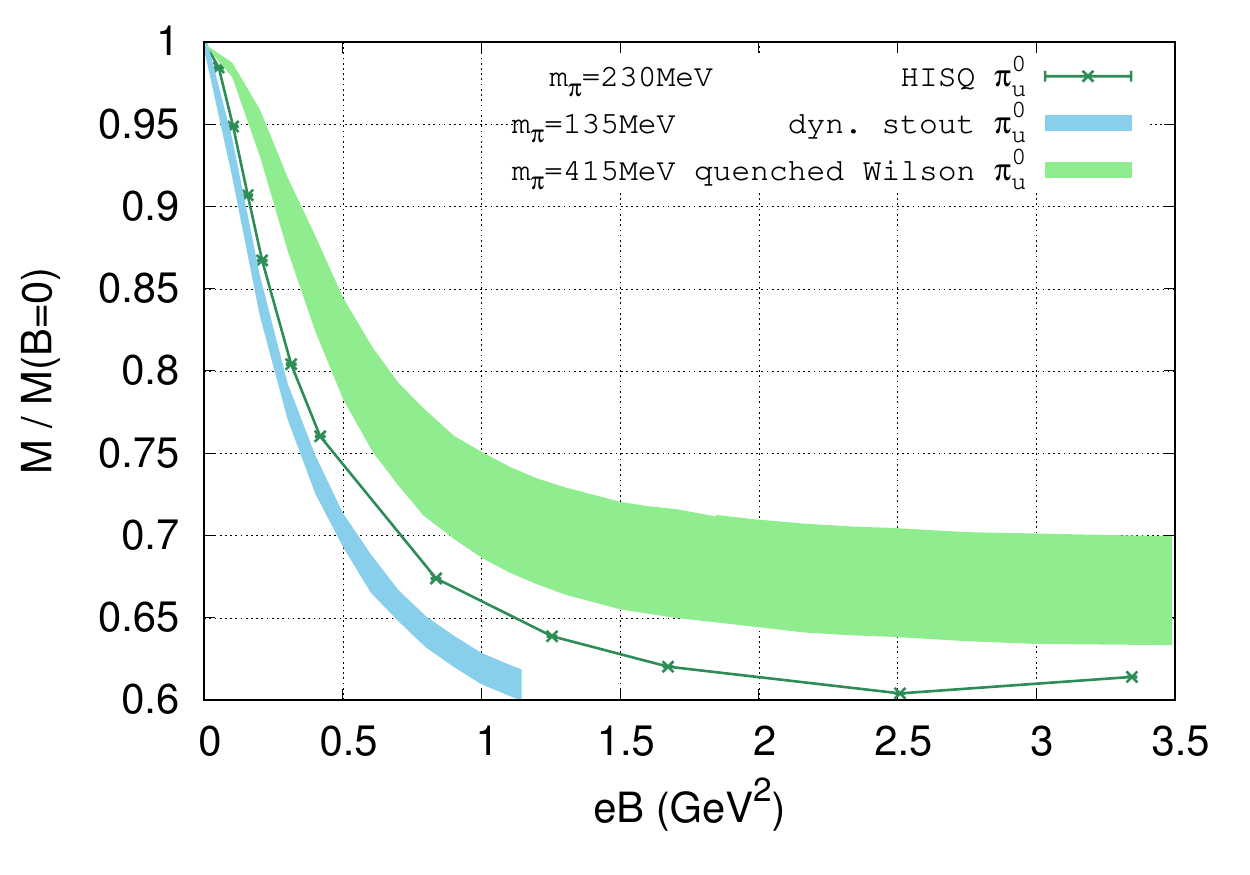}
\caption{Left: Masses of $\pi^0_u$, $\pi^0_d$, $K$, $\eta^0_s$ normalized by their zero field masses v.s. magnetic fields. Right: Comparison of $M_{\pi^0_u}$ v.s. $|eB|$ obtained in this work by using dynamic HISQ fermion at $m_\pi=230$ MeV (line) and dynamic stout fermion at $m_\pi=135$ MeV, quenched Wilson fermion at $m_\pi=415$ MeV (bands) \cite{Bali:2017ian}.}
\end{figure}

Firstly in the left plot of Fig. \ref{fig.neutral}, we see that  the normalized neutral pseudo-scalar particle $\pi^0_u, \pi^0_d$ masses ($M/M_{B=0}$) decrease as the external magnetic field grows and then saturate at very large values of $|eB|$ around 3 GeV$^2$. This is roughly consistent with previous studies \cite{luschevskaya2015magnetic,Bali:2017ian}, i.e. quenched result and dynamic stout result (see right plot of Fig. \ref{fig.neutral}). Due to the 30 $\sim$ 40\% mass reduction we observed, neutral mesons thus cannot be considered as point particles anymore in external magnetic field. Also the different magnitudes of the mass reduction between $\pi^0_u$ and $\pi^0_d$ may come from the different electric charges of up and down quarks which again indicates that the inner structure of meson should be taken into consideration. The magnetic field dependences of $K^0$, $\eta_s$ showed in the left of Fig. \ref{fig.neutral} have similar behavior as $\pi^0$. By comparing the normalized masses of $\pi^0_u$, $\pi^0_d$, $K^0$, $\eta_s$, it is obvious that the lighter hadrons are more affected by magnetic field. 

 \begin{figure}[htbp!]
\centering
\includegraphics[width=0.48\textwidth]{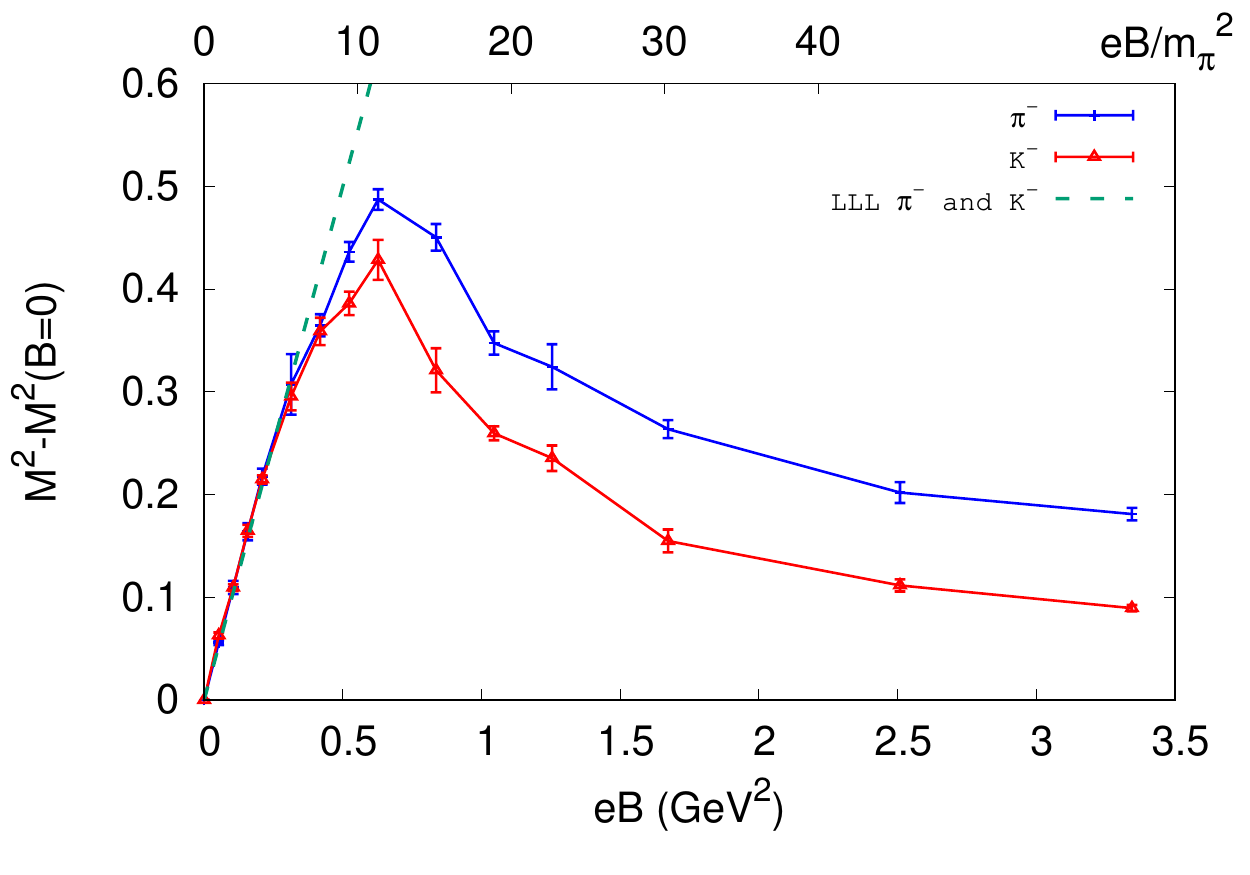}
\includegraphics[width=0.48\textwidth]{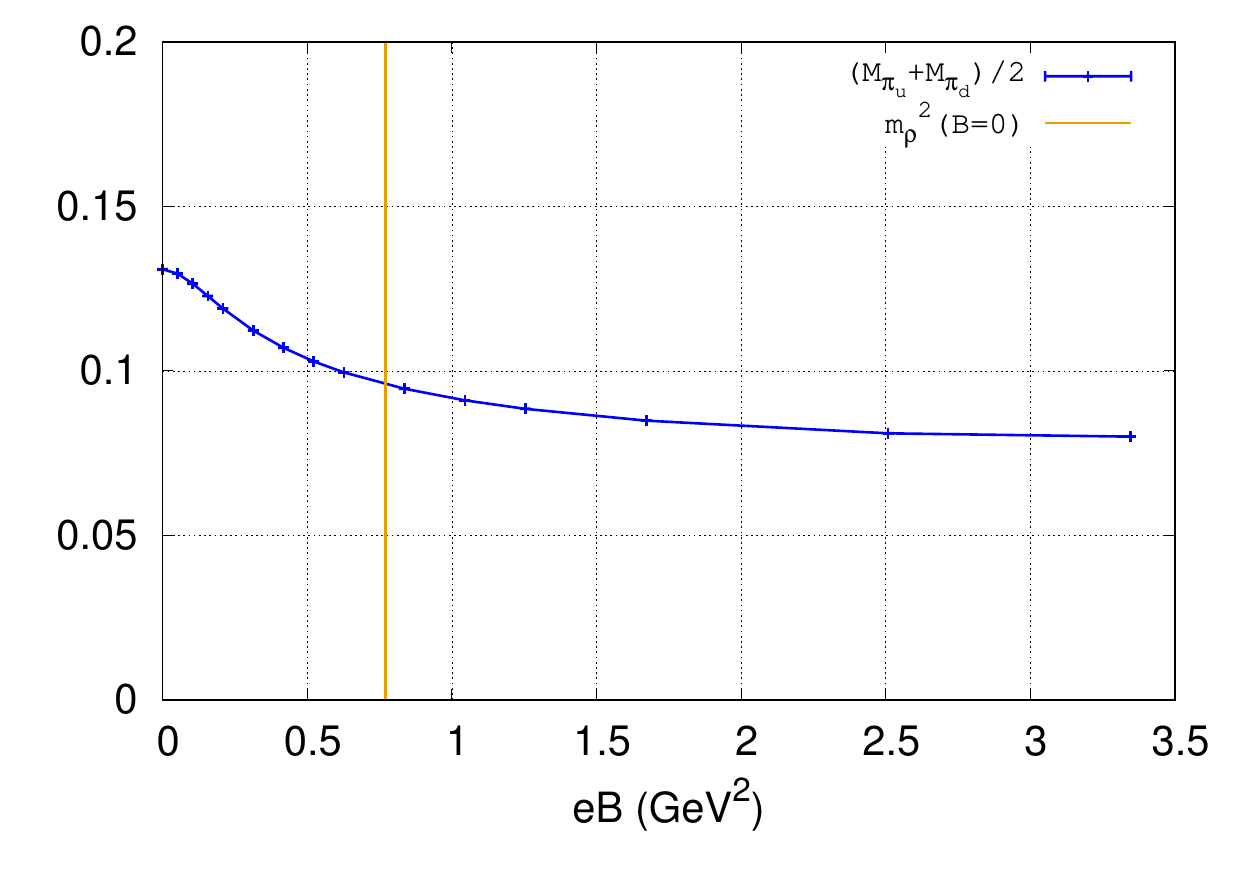}
\caption{Left: Mass square of $\pi^-$ and $K^-$ subtracted with their zero field mass square v.s. external  magnetic strength. The LLL mass $m^2= |eB|$ is characterised by the dashed line. Right: Average of $M_{\pi^0_u}$ and $M_{\pi^0_d}$ v.s. $|eB|$. The vertical line is the critical $|eB|$ value given by zero field $\rho$-meson mass.}\label{fig.charged}
\end{figure}

Next, we investigate the mass of charged pseudo-scalar particles. In small $|eB|$ regime, the energy of point-like meson can be described by mesonic Lowest Landau Level (LLL) through the following equation:
\begin{equation}\label{eq lll}
E_{n}^{2}=M_{\pi}^{2}+(2 n+1)|e B|+p_{z}^{2}.
\end{equation}
In the left plot of Fig. \ref{fig.charged}, we show the subtracted mass square ($M^2-M^2(B=0)$) of $\pi^-$ , $K^-$ as a function of $|eB|$ in comparison with the LLL approximation. The dashed line is simply $|eB|$ as with the lowest level ($n=0$) and zero momentum ($p_z=0$) in Eq.(\ref{eq lll}). Our results of $M_{\pi^-}$ , $M_{K^-}$ perfectly match with the LLL approximation in the small $|eB|$ regime ($|eB| \lesssim 0.3$ GeV). Beyond this regime the behavior cannot be explained by LLL approximation, and most likely inner structures of charged particles are probed by the magnetic field. More interestingly, we observe that $M_{\pi^-}$ , $M_{K^-}$ first increases and then decreases as magnetic field grows. We emphasize that this non-monotonous behavior is firstly observed and is different from previous quenched results. This non-monotonic behavior is also not seen in previous full QCD simulation probably as where eB is not sufficiently strong \cite{Bali:2011qj}.

In the right plot of Fig. \ref{fig.charged}, we show $|eB|$ dependence of  $(M_{\pi^0_u}+M_{\pi^0_d})/2$. And the critical value of magnetic field which equals the square of $\rho^\pm$ meson mass at $|eB|=0$ locates on the yellow vertical line i.e. $|eB|_c \approx 0.77$ GeV$^2$. According to QCD inequality $M_{\rho^{\pm}} \geq\left(M_{\pi_{u}^{0}}+M_{\pi_{d}^{0}}\right) / 2$, it is obvious that even the lower limit of  $\rho^\pm$ meson keeps nonzero at critical $|eB|$ value. It thus indicates that $\rho^\pm$ meson cannot condensate in the current window of external magnetic field.

\section{Conclusion}
\label{sec:conclusion}
We have investigated the mass of pseudo-scalar particles in presence of background magnetic fields by performing a full QCD simulation at zero temperature using HISQ fermions with $m_\pi\approx$230 MeV on 32$^3\times$96 lattices. We found that masses of neutral pseudo-scalar particles monotonously decrease as magnetic field grows and heavier neutral pseudo-scalars are less affected by magnetic fields. The decreasing behavior of $T_{pc}$ in $|eB|$ may be explained due to the decreasing of the mass of the lightest Goldstone boson $\pi^0$. In the case of charged pseudo-scalar particles, the non-monotonous behavior of their masses in $|eB|$ is firstly found. Mesonic LLL agrees quite well with our results at $|eB|\lesssim$ 0.3 GeV$^2$ and does not work any more at larger $eB$ where probably the inner structure of charged mesons is probed by the magnetic field.

Though vector channel particles' masses are hard to extract due to the mixing between pion and $\rho$-meson induced by magnetic field, the charged $\rho$-meson condensation can be investigated through neutral pion mass with the help of QCD inequalities. Since our neutral pion mass keeps nonzero in whole range of magnetic field strength we simulated, the existence of a superconducting phase is not favored.

\bibliographystyle{JHEP}
\bibliography{lat19_PoS_xdw}

\end{document}